\documentclass[12pt,a4paper,final]{iopart}

\usepackage{iopams}  
\usepackage{graphicx}
\usepackage[breaklinks=true,colorlinks=true,linkcolor=blue,urlcolor=blue,citecolor=blue]{hyperref}

\usepackage{setstack}
\usepackage[numbers,square,sort&compress]{natbib}

\usepackage{color}
\definecolor{bluecolor}{rgb}{0,0.,1.}

\definecolor{redcolor}{rgb}{.7,0.,0.}

\begin{document}

\title{Scaling laws and fluctuations in the statistics of word frequencies}

\author{Martin Gerlach and Eduardo G. Altmann}
\address{Max Planck Intsitute for the Physics of Complex Systems, 01187 Dresden, Germany}
\ead{{gerlach,edugalt}@pks.mpg.de}

\begin{abstract}
In this paper we combine statistical analysis of written texts and simple stochastic models to explain the appearance of scaling laws in the statistics of word frequencies. 
The average vocabulary of an ensemble of fixed-length texts is known to scale sub-linearly with the total number of words (Heaps' law).
Analyzing the fluctuations around this average in three large databases (Google-ngram, English Wikipedia, and a collection of scientific articles) we find that the standard deviation scales linearly with the
average (Taylor's law), in contrast to the prediction of decaying fluctuations obtained using simple sampling arguments.
We explain both scaling laws (Heaps and Taylor) by modeling the usage of words by a Poisson process with a fat-tailed distribution of word-frequencies (Zipf's law) and topic-dependent frequencies of
individual words (as in topic models).
Taking into account topical variations lead to quenched averages, turn the vocabulary size a non-self-averaging quantity, and explain the empirical observations.
For the numerous practical applications relying on estimations of vocabulary size, our results show that uncertainties remain large even for long texts.
We show how to account for these uncertainties in measurements of lexical richness of texts with different lengths.
\end{abstract}


\section{Introduction}
Fat-tailed distributions~\cite{clco+09,mi04,ne05}, allometric scaling~\cite{West1997,Bettencourt2007a}, and fluctuation scaling~\cite{TAYLOR1961,DeMenezes2004,Eisler2008a} are the most prominent examples of scaling laws appearing in complex systems.
Statistics of words in written texts provide some of the best studied examples: the  fat-tailed  distribution of word frequencies (Zipf's law)~\cite{Zipf1949} and the sublinear growth (as in allometric scalings) of the number of distinct words as a   function of database size (Heaps' law)~\cite{HERDAN1958,Heaps1978}.
The connection between these two scalings is known at least since Mandelbrot~\cite{Mandelbrot1961}, and has been further investigated in
recent years~\cite{VANLEIJENHORST2005,Zanette2005,sefl+09}, especially for large databases~\cite{Williams2005}, finite text sizes~\cite{beco+09,Lu2010}, and more general distributions~\cite{Gerlach2012,Font-Clos2013}. 
In this paper we report the existence of a third type of scaling in the statistics of words:  fluctuation scaling. It appears when investigating  the   
fluctuations around the Heaps' law, i.e., the variance of the vocabulary over different texts  of the same size scales with the average. 
We show that this scaling results from topical aspects of written text which are ignored in the usual connection between Zipf's and Heaps' law.

The importance of looking at the fluctuations around Heaps' law is that this law is used in different applications~\cite{Manning2008}, e.g., 
(i) to optimize the memory allocation in inverse indexing algorithms~\cite{Baeza-Yates2000}; (ii) to estimate the
vocabulary of a language~\cite{mish+11,Klein2013}; (iii) to compare the vocabulary
richness of documents with different lengths~\cite{Wimmer1999,Baayen2001,Yasseri2012a}.  
Beyond linguistic applications, scalings of the number of unique items as a function of
database size similar to Heaps' law have been observed in
other domains, e.g. the species-area relationship in
ecology~\cite{Brainerd1982,GarciaMartin2006}, collaborative tagging~\cite{Cattuto2009},
network growth~\cite{Krapivsky2013}, and in the statistics of chess
moves~\cite{Perotti2013}. These scaling laws have been analyzed from the general viewpoint
of innovation dynamics~\cite{Tria2013} and sampling problems~\cite{Gnedin2007}.
Our results allow for the quantification of uncertainties in the estimation of these
scaling laws and lead to a rethinking of the statistical significance of previous
findings.

We use as databases three different collections of texts: i) all articles of the English Wikipedia~\cite{Wikimedia}, ii) all articles published in the journal PlosOne~\cite{Plosone}, and  iii) the Google-ngram database~\cite{mish+11}, a collection of books published in $1520-2008$ (each year is treated as a separate document). See  \ref{sec.app.data} for details on the data.

The manuscript is divided as follows. 
Section~\ref{sec.data} reports our empirical findings with focus on the deviations from a Poisson null model. 
Section~\ref{sec.topic} shows how these deviations can be explained by including topicality, which plays the role of a quenched disorder and leads to a non-self averaging process. 
The consequences of our findings to applications, e.g. vocabulary richness, are discussed in Sec.~\ref{sec.applications}. 
Finally, Sec.~\ref{sec.discussion} summarizes our main results.

\section{Empirical Scaling Laws}\label{sec.data.scaling}\label{sec.data}
The most-prominent scaling in language is Zipf's law~\cite{Zipf1949} which states that the frequency, $F$, of the $r$-th most frequent word (i.e., the fraction of times it occurs in the database) scales as
\begin{equation}\label{eq.Zipf}
 F_r \propto r^{-\alpha}\,\,\mathrm{for}\,\, r\gg 1.
\end{equation}
Another well-studied scaling in language concerns the vocabulary growth and is known as Heaps' law~\cite{HERDAN1958,Heaps1978}. It states that the number of different words, $N$, scales sublinearly with the total number of words, $M$, i.e.
\begin{equation}\label{eq.Heaps}
 N(M) \propto M^{\lambda} \,\,\mathrm{for}\,\, M\gg 1,
\end{equation}
with $0< \lambda < 1$.
As a third case, we consider here the problem of the vocabulary growth for an ensemble of texts, and study the scaling of fluctuations by looking at the relation between the standard deviation, $\sigma(M)=\sqrt{\mathbb{V} \left[ N(M) \right]}$, and the mean value,
$\mu(M)=\mathbb{E} \left[ N(M) \right]$, computed over the ensemble of texts with the same textlength $M$. 
In other systems, Taylor's law~\cite{TAYLOR1961} 
\begin{equation}\label{eq.Taylor}
 \sigma(M) \propto \mu(M)^{\beta} \,\,\mathrm{for}\,\, \mu(M) \gg 1
\end{equation}
with $ 1/2 \leq \beta \leq 1 $ is typically observed~\cite{Eisler2008a}.

The connection between scalings~(\ref{eq.Zipf}) and~(\ref{eq.Heaps})  (Zipf's and Heaps'
law) can be revealed assuming the usage of each word $r$ is governed by an independent Poisson process with a given frequency $F_r$. 
In this description, the number of different words, $N$, becomes a stochastic variable 
for which we can calculate the expectation value $\mathbb{E} \left[ N(M) \right]$ and the
variance $\mathbb{V} \left[ N(M) \right]$ over the realizations of the Poisson
process (see \ref{sec.app.poissonnull} for details) 
\begin{eqnarray}
  \mathbb{E} \left[ N(M) \right] &\equiv \mu(M) &=  \sum_r 1- e^{-M F_r} , \label{eq.ze.mean}\\
  \mathbb{V} \left[ N(M) \right] &\equiv \sigma(M)^2 & \equiv \mathbb{E} \left[ N(M)^2 \right] -\mathbb{E}
  \left[ N(M) \right]^2= \sum_r e^{-M F_r} - e^{-2 M F_r}.\label{eq.ze.sdev}
\end{eqnarray}
Assuming Zipf's law~(\ref{eq.Zipf}), for $M \gg 1$ we recover Heaps' law~(\ref{eq.Heaps}), i.e. $\mathbb{E} \left[ N(M) \right] \propto M^{\lambda}$, with a simple relation between the scaling exponents $\alpha = \lambda^{-1}$~\cite{Eliazar2011} and Taylor's law~(\ref{eq.Taylor}) with $\beta = 1/2$.

\begin{figure*}[!bt]
\centering
\includegraphics[width=0.95\columnwidth]{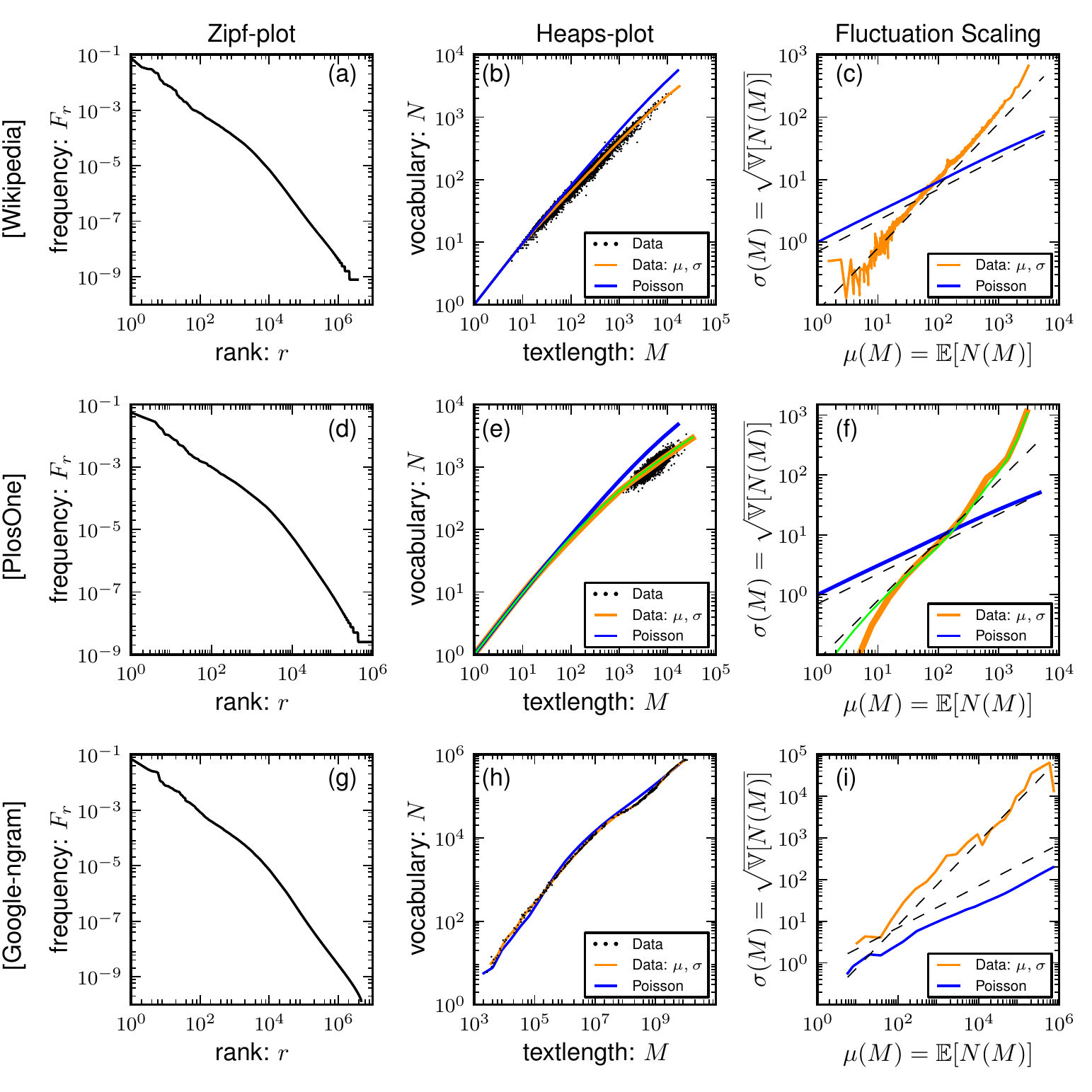}
\caption{
Scaling of Zipf's law~(\ref{eq.Zipf}), Heaps' law~(\ref{eq.Heaps}), and fluctuation
scaling~(\ref{eq.Taylor}). 
Each row corresponds to one of the three databases used in our work.
(a,d,g) Zipf's law: Rank-frequency distribution $F_r$ considering the full database (the
double power-law nature of the curves is apparent~\cite{Gerlach2012}).
(b,e,h) Heaps' law: the number of different words, $N$, as a function of textlength, $M$,
for each individual article in the corresponding database (black dots). 
(c,f,i) Fluctuation scaling: standard deviation, $\sigma(M)$, as a function of the mean, $\mu(M)$, for the vocabulary $N(M)$ conditioned on the textlength $M$. 
Poisson (blue-solid) shows the expectation from the Poisson null model, Eqs.~(\ref{eq.ze.mean},\ref{eq.ze.sdev}), assuming the empirical rank-frequency distribution from (a,d,g), respectively. 
(Data: $\mu,\sigma$) (yellow-solid) shows the mean, $\mu(M)$, and standard deviation, $\sigma(M)$, of the data $N(M)$ within a running window in $M$ (see \ref{sec.app.data} for the details on the procedure).
Additionally, (e,f) show the results (Data: $\mu,\sigma$) obtained shuffling the word order for each individual article (thin green-solid).
The fact that this curve is indistinguishable from the original curve shows that the results are not due to temporal correlations within the text.
For comparison, we show in (c,f,i) the scalings $\sigma(M) \propto \mu(M)^{1/2}$ and $\sigma(M) \propto \mu(M)$ (dashed).
}
\label{fig.1}
\end{figure*}

In Fig.~\ref{fig.1} we show empirical data of real texts for the scaling
relations~(\ref{eq.Zipf})-(\ref{eq.Taylor}) and compare them with predictions from the
Poisson null model in Eqs.~(\ref{eq.ze.mean},\ref{eq.ze.sdev}).  
The Poisson null model correctly elucidates the connection between the scaling exponents
in Zipf's and Heaps' law, but it suffers from two severe drawbacks.
First, it is of limited use for a quantitative prediction of the vocabulary size for individual articles as it systematically overestimates its magnitude, see Fig.~\ref{fig.1}(b,e,h).
Second, it dramatically underestimates the expected fluctuations of the vocabulary size yielding a qualitatively different behavior in the fluctuation scaling:
whereas the Poisson null model yields an exponent $\beta \approx 1/2$ expected from
central-limit-theorem-like convergence~\cite{Eisler2008a}, the three empirical data [Fig.~\ref{fig.1}(c,f,i)] exhibit a
scaling with $\beta \approx 1$.
This implies that relative fluctuations of $N$ around its mean value $\mu$ for fixed $M$ do not decrease with larger text size (the
vocabulary growth, $N(M)$, is a non-self-averaging quantity) and remain of the order of the expected value.
Indeed, we find that in all three databases
\begin{equation}\label{eq.taylor.quant}
 \sigma(M) \approx 0.1 \mu(M).
\end{equation}

Instead of looking at a single value $(N,M)$ for each document, as described above, an alternative approach is to count the number of different words, $N$, in the first $M$ words of the document. 
This leads to a curve $N(M)$ for $M=1,2,\ldots, M_{\mathrm{max}}$, where $M_{\mathrm{max}}$ is the length of the document. 
This alternative approach was employed in Fig.~\ref{fig.1}(e,f) and leads to results equivalent to the ones obtained using single values $(N,M)$, i.e. the $\mu(M)$ and $\sigma(M)$ obtained over different texts lead to identical Heaps' and Taylor's laws.
In Fig.~\ref{fig.1}(f) we show that anomalous fluctuation scaling in the vocabulary growth is preserved if shuffling the word order of individual texts.
This illustrates that in contrast to usual explanations of fluctuation scaling in terms of long-range correlations in time-series~\cite{Eisler2008a}, here, the observed deviations from the Poisson null model are mainly due to fluctuations across different texts.

In the following, we argue that these observations can be accounted for by considering topical aspects of written language, i.e. instead of treating word-frequencies as fixed,
we will consider them to be topic-dependent ($F_r \mapsto F_r(\mathrm{topic})$).

\section{Topicality in the Vocabulary Growth}\label{sec.topic}
\subsection{Topicality}\label{sec.topic.topicality}
\begin{figure*}[!bt]
\centering
\includegraphics[width=1\columnwidth]{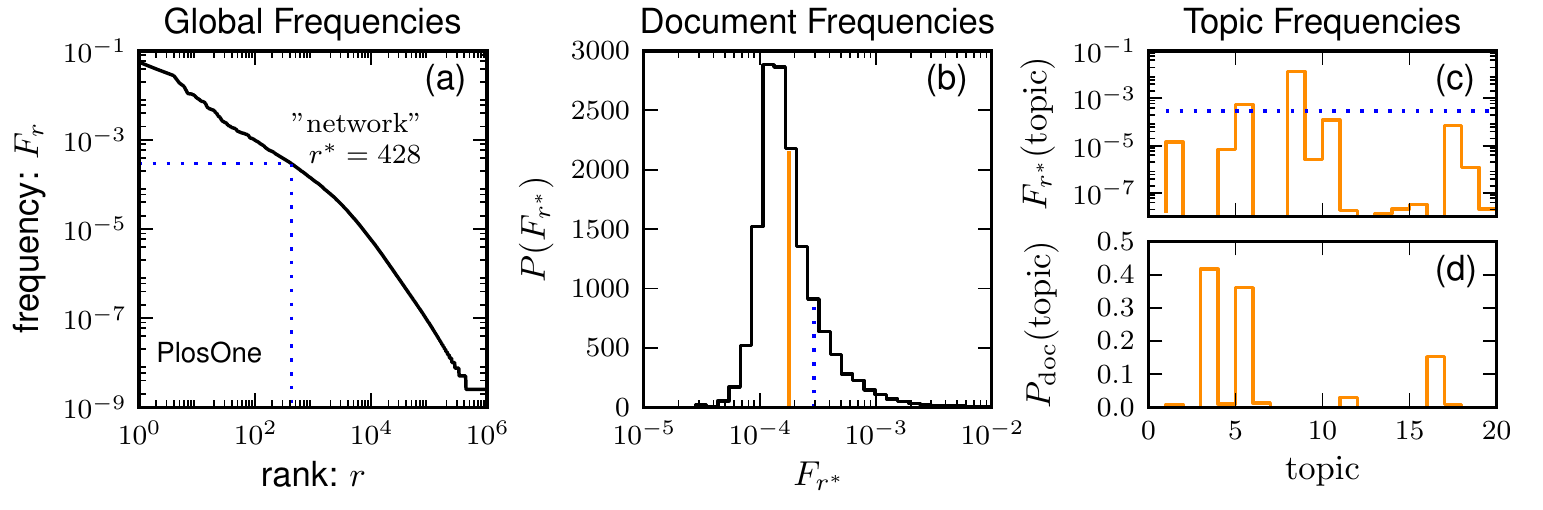}
\caption{Variation of frequencies due to topicality in the PlosOne database. 
(a) Rank-frequency distribution considering the complete database. The word ``network" (dotted line) has $F_{r^*=428} \approx 2.9 \times 10^{-4}$.
(b) Distribution $P(F_{r^*})$ of the local frequency $F_{r^*}$ obtained from each article separately for the word ``network" with the global frequency from (a) (dotted).
(c) Topic-dependent frequencies $F_{r^*}( \mathrm{topic})$ inferred from LDA with $T=20$ topics for the word ``network" with global frequency from (a) as comparison (dotted).
(d) One realization for the topic composition of a single document, $P_{\mathrm{doc}}(\mathrm{topics})$, drawn from a Dirichlet distribution. For this realization, the effective frequency is $F_{r,\mathrm{doc}}=\sum_{t=1}^T P_{\mathrm{doc}}(t) F_r(t) \approx 2.0 \times 10^{-4}$ and is shown in (b) (solid).
}
\label{fig.2}
\end{figure*}

The frequency of an individual word varies significantly across different texts  meaning
that its usage cannot be described alone by a single global
frequency~\cite{Church1995,Montemurro2010,Altmann2011}. 
For example, consider the usage of the (topical) word ``network'' in all articles
published in the journal PlosOne. It has an overall rank $r^*=428$ and a global frequency,
$F_{r^*=428} \approx 2.9 \times 10^{-4}$, see Fig.~\ref{fig.2}(a). The local frequency
obtained from each article separately varies over more than one decade, see Fig.~\ref{fig.2}(b).
Note that, although in this case the local rank-ordering differs from document to document, the index $r$ still refers to the globally determined rank and is used as a unique label for each word. 

One popular approach to account for the heterogeneity in the usage of single words are
\textit{topic models}~\cite{Blei2012}. The basic idea is that the variability across
  different documents can be explained by the existence of (a smaller number of) topics.
In the framework of a generative model it assumes i) that individual documents are
composed of a mixture of topics (indexed by $t=1,..,T$), with each topic represented in an individual document by the probabilities $P_{\mathrm{doc}}(\mathrm{topic}=t)$; and ii) that the frequency of each word is topic-dependent, i.e. $ F_r(\mathrm{topic}=t)$, which leads to a different effective frequency in each document, $F_{r,\mathrm{doc}}=\sum_{t=1}^T P_{\mathrm{doc}}(t) F_r(t)$.
One particularly popular variant of topic models is Latent Dirichlet Allocation (LDA)~\cite{Blei2003}, which assumes that the topic composition $P_{\mathrm{doc}}(\mathrm{topic})$ of each document is drawn from a Dirichlet distribution, $P_{\mathrm{Dir}}$, such that only few topics contribute substantially to each document.
  Given a database of documents, LDA infers the topic-dependent frequencies, $F_r(\mathrm{topic})$, from numerical maximization of the posterior likelihood of the generative model~\cite{gensim}.
As an illustration, in Fig.~\ref{fig.2}(c) we show $F_{r^*}(\mathrm{topic})$ obtained using LDA for the word
``network" in the PlosOne database. As expected from a meaningful topic model, we see that the conditional frequencies vary over many
orders of magnitude, and that the global frequency $F_{r^*}$ is governed by few topics. 
The advantage of LDA is that, instead of measuring the distribution of frequencies of each individual word (or $2$-point distributions for assessing correlations) over different documents, it estimates the frequency of individual words for a finite (and small) number of topics. 
In combination with the generative model (e.g., drawing $P_{\mathrm{doc}}(\mathrm{topic})$ from a Dirichlet distribution), this not only yields a more compact description of topicality by dramatically reducing the number of parameters, but also allows for an easy extrapolation to unseen texts from a small training sample~\cite{Blei2003}.

\subsection{General treatment}\label{sec.topic.gen}
In this section we show how topicality can be included in the analysis of the vocabulary
growth. 
The simplest approach is to consider again that the usage of each word is governed by Poisson processes, but this time to consider that frequencies are not fixed but
are themselves random variables that vary across texts.

In this setting, the random variable representing the vocabulary size, $N$, for a text of length $M$ can be written as
\begin{equation}
 N(M) = \sum_r I\left[ n_r (M, F_r)  \right],
\end{equation}
in which $n_r$ is the integer number of times the word $r$ occurs in a Poisson
process of length $M$ with frequency $F_r$ and $I[x]$ is an indicator-type function,
i.e. $I[x=0]=0$ and $I[x\ge1]=1$ .
The calculation of the expectation value now consists of two parts: i) the average over
realizations $i$ of the Poisson processes $n^{(i)}_r(M,F^{(j)}_r)$ for a given realization
$j$ of the set of frequencies $F^{(j)}_r$; and ii) the average over all possible realizations $j$ of the sets of frequencies $F^{(j)}_r$ (which vary due to topicality).
In this framework expectation values correspond to quenched averages (denoted by subscript
$q$) 
\begin{equation} \label{eq.zegen.mean}
  \mathbb{E}_q \left[ N(M) \right] 
  = \left\langle N(M)^{(i,j)} \right\rangle_{i,j} 
  = \sum_r \left\langle I\left[ n^{(i)}_r (M, F^{(j)}_r)  \right] \right\rangle_{i,j} 
  = \sum_r  1- \langle e^{-MF_r^{(j)}} \rangle_{j},  
\end{equation}
where we used 
\begin{equation}\label{eq.ze.poiss}
 \left\langle I\left[n^{(i)}_r (M, F^{(j)}_r)  \right] \right\rangle_i  = 1 - P(n_r=0;M,F_r^{(j)}) = 1 - e^{-MF_r^{(j)}}.
\end{equation}
The last equation corresponds to the probability of word $r$ not occurring for a Poisson
process of duration $M$ with frequency $F_r^{(j)}$, as in Eq.~(\ref{eq.ze.mean}). For
simplicity, hereafter $\left\langle \ldots \right\rangle\equiv \left\langle \ldots \right\rangle_j$ (the average over realizations of sets of frequencies $F_r^{(j)}$).

Using the inequality between arithmetic and geometric mean
\begin{equation}\label{eq.mean.inequ}
 e^{\left\langle \ln x \right\rangle}= \left\langle x \right\rangle_{\mathrm{geometric}}  \leq   \left\langle x \right\rangle_{\mathrm{arithmetic}} = \left\langle e^{ \ln x } \right\rangle,
\end{equation}
we obtain that
\begin{equation}
\mathbb{E}_q \left[ N(M) \right] =\sum_r 1 - \left\langle e^{-MF_r} \right\rangle \leq \sum_r 1 -  e^{-M \left\langle F_r \right\rangle} \equiv \mathbb{E}_a \left[ N(M) \right].
\end{equation}
The right hand side corresponds to the result of the Poisson null model (with fixed $F_r=\langle F_r
\rangle$), see Eq.~(\ref{eq.ze.mean}), and can be interpreted as an annealed average (denoted by subscript $a$).
This implies that the heterogeneous dissemination of words across different texts leads to
a reduction of the expected size of the vocabulary, in agreement with the first deviation
of the Poisson null model reported in Fig.~\ref{fig.1}(b,e,h).

For the quenched variance we obtain (see \ref{sec.app.N2})
\begin{eqnarray}
 \mathbb{V}_q \left[ N(M) \right] 
  &\equiv& \mathbb{E}_q \left[ N(M)^2 \right] - \mathbb{E}_q \left[ N(M) \right]^2\\
  &=&\sum_r \left\langle e^{-MF_r} \right\rangle - \left\langle e^{-MF_r} \right\rangle^2 + \sum_r \sum_{r'\neq r} \mathrm{Cov}[e^{-MF_{r}},e^{-MF_{r'}} ] \label{eq.zegen.sdev}
\end{eqnarray}
where $\mathrm{Cov}[e^{-MF_{r}},e^{-MF_{r'}}] \equiv \left\langle e^{-MF_{r}} e^{-MF_{r'}} \right\rangle - \left\langle e^{-MF_{r}} \right\rangle \left\langle e^{-MF_{r'}} \right\rangle$.
Comparing to the Poisson case in Eq.~(\ref{eq.ze.sdev}), we see that the quenched
average yields an additional term containing the correlations of different words. In
general, this term does not vanish and is responsible for the anomalous fluctuation
scaling with $\beta=1$ observed in real text, explaining the second deviation from the
Poisson null model reported in Fig.~\ref{fig.1}(c,f,i).

\subsection{Specific ensembles}\label{sec.topic.spec}
In this section we compute the general results from Eqs.~(\ref{eq.zegen.mean},\ref{eq.zegen.sdev}) for particular ensembles of frequencies $F_r^{(j)}$ and compare them to the empirical results.
In the absence of a generally accepted parametric formulation of such an ensemble, we propose two nonparametric approaches explained in the following.

In the first approach we construct the ensemble $F_r^{(j)}$ directly from the collection of documents, i.e. the frequency $F_r^{(j)}$ corresponds to the frequency of word $r$ in document $j$, such that 
\begin{equation}\label{eq.ens.direct}
 \left\langle e^{-MF_r} \right\rangle = \frac{1}{D}\sum_{j=1}^D e^{-MF_r^{(j)}},
\end{equation}
where $D$ is the number of documents in the data, see Fig.~\ref{fig.2}(b).

In the second approach we construct the ensemble from the LDA topic model~\cite{Blei2003}, 
in which $F_r^{(j)} = F_r(\mathrm{topic}=j)$ corresponds to the frequency of word $r$ conditional on the topic $j=1...T$, see Fig.~\ref{fig.2}(c+d).
In this particular formulation each document is assumed to consist of a composition of topics, $P_{\mathrm{doc}}(\mathrm{topic})$, which is drawn from a Dirichlet distribution, such that we get for the quenched average
\begin{equation}\label{eq.ens.lda}
 \left\langle e^{-MF_r} \right\rangle = \int \mathrm{d} \theta P_{\mathrm{Dir}}(\theta | \alpha) e^{-MF_r (\theta)},
\end{equation}
in which $\theta=(\theta_1,...,\theta_T)$ are the probabilities of each topic, $F_r (\theta) = \sum_{j=1}^T \theta_j F_r(\mathrm{topic}=j)$, and the integral is over a $T$-dimensional Dirichlet-distribution $P_{\mathrm{Dir}}(\theta | \alpha)$ with concentration parameter $\alpha$.
We infer the $F_r(\mathrm{topic})$ using Gensim~\cite{gensim} for LDA with $T=100$ topics.

\begin{figure*}[!bt]
\centering
\includegraphics[width=1.0\columnwidth]{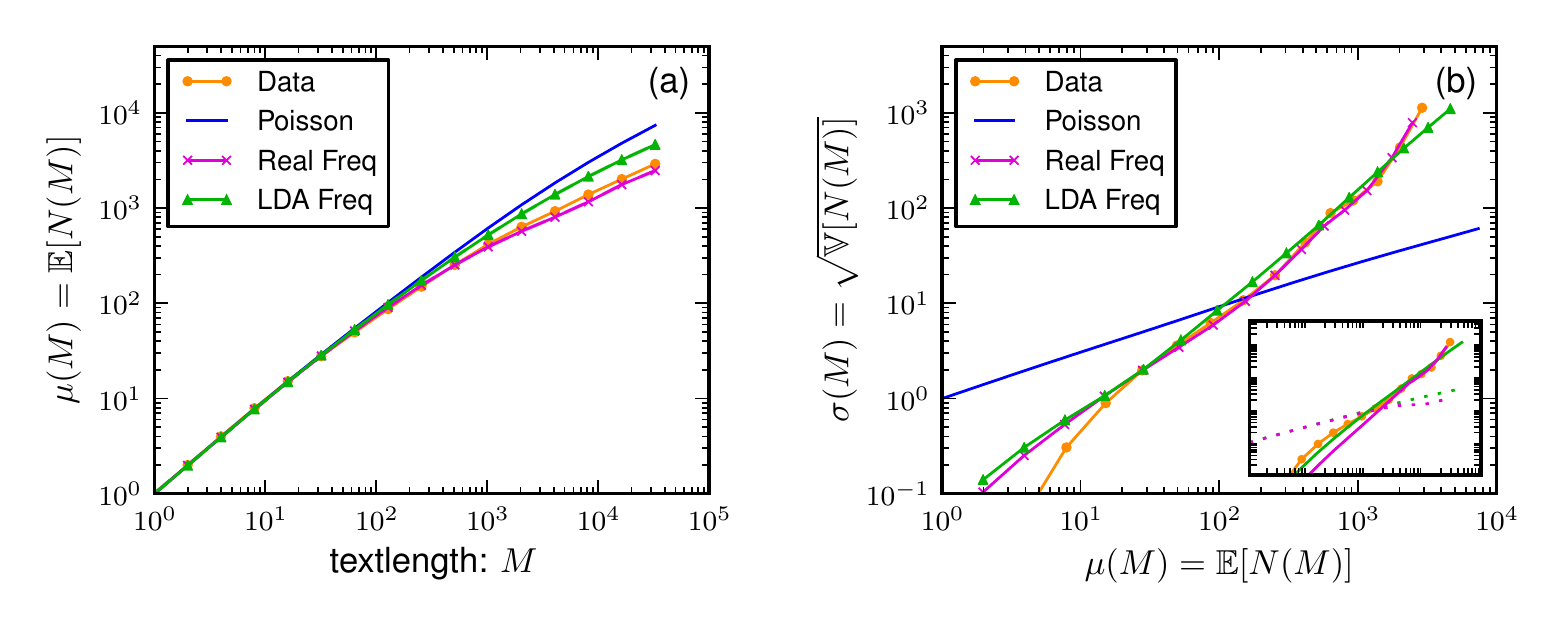}
\caption{Vocabulary growth for specific topic models. (a) Average vocabulary growth and
  (b) fluctuation scaling in the PlosOne database (Data) and in the calculations from Eqs.~(\ref{eq.zegen.mean},\ref{eq.zegen.sdev}) for the two topic models based on the measured frequencies in individual articles (Real Freq) and on LDA (LDA Freq), compare Eqs.~(\ref{eq.ens.direct},\ref{eq.ens.lda}). 
  For comparison we show the results from the Poisson null model (Poisson), Eqs.~(\ref{eq.ze.mean},\ref{eq.ze.sdev}), which do not take into account topicality.
  The inset in (b)  (same scale as main figure) shows the individual contributions to the fluctuations in Eq.~(\ref{eq.zegen.sdev}): $\sum_r \left\langle e^{-MF_r} \right\rangle - \left\langle e^{-2MF_r} \right\rangle$ (dotted) and $\sum_r \sum_{r'\neq r} \mathrm{Cov}[e^{-MF_r},e^{-MF_{r'}}]$ (solid), illustrating that correlations between different words lead to anomalous fluctuation scaling. 
  The solid lines for LDA-Freq and Real Freq in (b) show the calculations of the corresponding topic models replacing the Poisson by multinomial usage in the derivation of Eqs.~(\ref{eq.zegen.mean},\ref{eq.zegen.sdev}) in order to avoid finite-size effects for $\mu(M)<100$.
  }
\label{fig.3}
\end{figure*}

The results from both approaches are compared to the PlosOne database in Fig.~\ref{fig.3}. 
Fig.~\ref{fig.3}(a) shows that both methods lead to a reduction in the mean number of different words. 
Whereas the direct ensemble, Eq.~(\ref{eq.ens.direct}), almost perfectly matches the curve of the data, the LDA-ensemble, Eq.~(\ref{eq.ens.lda}), still overestimates the mean number of different words in the data. 
This is not surprising since due to the fewer number of topics (when compared to the
number of documents) it constitutes a much more coarse-grained description than the direct ensemble. Additionally, the LDA-ensemble relies on a number of ad-hoc assumptions, e.g. the Dirichlet-distribution in Eq.~(\ref{eq.ens.lda}) or the particular choice of parameters in the inference algorithm which were not optimized here.
More importantly, both methods correctly account for the anomalous fluctuation scaling
with $\beta=1$ observed in the real data, see Fig.~\ref{fig.3}(b) and even yield a similar proportionality factor in the quantitative agreement with the data.
The comparison of the individual contributions to the fluctuations, Eq.~(\ref{eq.zegen.sdev}), shown in the inset of Fig.~\ref{fig.3}(b) shows that the anomalous fluctuation scaling is due to correlations in the co-occurrence of different words (contained in the term $\mathrm{Cov}[e^{-MF_r},e^{-MF_{r'}}]$).

\section{Applications}\label{sec.applications}

\subsection{Adding texts}

In thermodynamic terms, Heaps' law (as other allometric scalings) implies that the
  vocabulary size is neither extensive nor intensive ($N(M) < N(2M) < 2 N(M)$, also for
  $M\rightarrow \infty$).  While this can be seen as a direct consequence of Zipf's law,
  our results show that Heaps' law depends also sensitively on the
  fluctuations of the frequency of specific words across different documents. To illustrate
this, consider the problem of doubling the size of a text of size $M$. This can be done
either by simply extending the size of the same text up to size $2M$ (denoted by $M'=2 \cdot
M$) or by concatenating another text of size $M$ (denoted by $M'=2 \times M$). 
The Poisson model (fixed frequency or annealed average) predicts the same expected vocabulary for both procedures
\begin{equation}
\mathbb{E}_a[N(2 \cdot M)] = \mathbb{E}_a[N(2 \times M)] = \sum_r 1-e^{-2M  \left\langle F_r \right\rangle}.
\end{equation}
Taking fluctuations of individual frequencies across documents (quenched average) into account yields (see \ref{sec.app.add} for details):
\begin{equation}
\mathbb{E}_q[N(2 \cdot M)] = \sum_r 1- \left\langle e^{-2 M F_r } \right\rangle \; \mathrm{ and } \; \mathbb{E}_q[N(2 \times M)] = \sum_r 1- \left\langle e^{-M	 F_r } \right\rangle^2.
\end{equation}
Using Eq.~(\ref{eq.mean.inequ}) and the fact that $ \left\langle x^2 \right\rangle \geq \left\langle x \right\rangle^2$, we obtain the following general result
\begin{equation}
 \mathbb{E}_q[N(2\cdot M)] \leq \mathbb{E}_q[N(2 \times M)] \leq \mathbb{E}_a[N(2 \times M)] = \mathbb{E}_a[N(2 \cdot M)].
\end{equation}
This is consistent with the intuition that the concatenation of different texts (e.g., on different topics) leads to larger vocabulary than
a single longer text. The calculations above remain true if the text is extended by a factor $k$ (instead of $2$), even for $k\rightarrow
\infty$. 

The fluctuations around the mean show a more interesting behavior, as revealed by repeating
the computations above for the variance. We consider the case of $k$ texts each of length
$M$, such that $M'=k \times M$, and focus on the terms containing correlations between different
words shown to be responsible for the anomalous fluctuation scaling (see \ref{sec.app.add} for
details):
\begin{equation}
 \mathbb{V}_q[N(k \times M)] \sim  \sum_{r,r'} \left\langle e^{-M F_r} e^{-M F_{r'}} \right\rangle^k -  \left\langle e^{-M F_r} \right\rangle^k \left\langle e^{-M F_{r'}} \right\rangle^k.
\end{equation}
The individual terms can be written as
\begin{eqnarray}
 \sum_{r,r'} \langle e^{-M F_r} e^{-M F_{r'}} \rangle^k  &=& \langle [ \sum_r e^{-Mk \bar{F}_r^{(k)} } ]^2 \rangle_{j_1,...,j_k},\\
 \sum_{r,r'} \langle e^{-M F_r} \rangle^k \langle e^{-M F_{r'}} \rangle^k &=& [ \langle  \sum_r e^{-Mk \bar{F}_r^{(k)} } \rangle_{j_1,...,j_k} ]^2,
\end{eqnarray}
in which $\left\langle \cdot \right\rangle_{j_1,...,j_k}$ denotes the averaging over the realizations $(j_1,...,j_k)$ of frequencies $F_r^{(j_i)}$ in each single text $i=1,...,k$ and $\bar{F}_r^{(k)} = \frac{1}{k}\sum_{i=1}^k F_r^{(j_i)}$ is the $k$-sample average frequency based on the realizations $(j_1,...,j_k)$.
In the limit $k \rightarrow \infty$: $\bar{F}_r^{(k)} \longrightarrow \left\langle F_r \right\rangle$ such that 
\begin{equation}
  \sum_{r,r'} \left\langle e^{-M F_r} e^{-M F_{r'}}\right\rangle^k - \left\langle e^{-M F_r}\right\rangle^k \left\langle e^{-M F_{r'}} \right\rangle^k \rightarrow 0
\end{equation}
for $k \rightarrow \infty$.
This implies that for $k\gg 1$ (adding many different texts) the fluctuations in the
vocabulary across documents (and therefore the correlations between different words) vanish and
normal fluctuation scaling ($\beta=1/2$) is recovered. This prediction can be tested in data. Starting from a collection of documents, we create a new collection by
concatenating $k$ randomly selected documents (each document is used once). We then
compute for each concatenated document the number of distinct words $N$ up to size $M$ for increasing $M$, $\mathbb{E}[N(M)]$, and $\mathbb{V}[N(M)]$. We observe a
transition of the exponent $\beta$ in the fluctuation scaling, Eq.~(\ref{eq.Taylor}), from
$\beta \approx 1 \longrightarrow \beta \approx 1/2$.

\subsection{Vocabulary Richness}
When measuring vocabulary richness we want a measure which is robust to different text sizes.
The traditional approach is to use Herdan's C, i.e. $C=\log N / \log M$~\cite{Wimmer1999,Baayen2001,Yasseri2012a}. 
While quite effective for rough estimations, this approach has several problems. 
One obvious one is that it does not incorporate any deviations from the original Heaps' law (e.g., the double scaling regime~\cite{Gerlach2012}). 
More seriously, it does not provide any estimation of the statistical significance or expected fluctuations of the measure.
For instance, if two values are measured for different texts one can not determine whether one is significantly larger than the other. 
Our approach is to compare observations with the fluctuations expected from models in the spirit of Sec.~\ref{sec.topic.gen}.

The computation of statistical significance requires an estimation of the probability of finding $N$ different words in a text of length $M$, $P(N | M)$, which can be obtained from a given generative model (e.g., as presented in Sec.~\ref{sec.topic}).
For a text with $(N^*,M^*)$ we compute the percentile $P(N>N^*| M^*)$, which allows for a ranking of texts with different sizes such that the smaller the percentile, the richer the vocabulary. 
An estimation of the significance of the difference in the vocabulary can then be obtained by comparison of the different percentile.

For the sake of simplicity, we illustrate this general approach by approximating $P(N | M)$ by a Gaussian distribution. 
In this case, the percentile are determined by the mean, $\mu(M)=\mathbb{E}[N(M)]$,  and the variance, $\sigma(M)=\sqrt{\mathbb{V}[N(M)]}$, in terms of the z-score
\begin{equation}\label{eq.appl.z1}
 z_{(N,M)}= \frac{N - \mu(M)}{\sigma(M)},
\end{equation}
which shows how much the measured value $(N,M)$ deviates from the expected value $\mu(M)$ in units of standard deviations ($z_{(N,M)}$ follows a standard normal distribution: $z \overset{d}{\sim} \mathcal{N}(0,1)$). 
If we take into account our quantitative result on fluctuation scaling in the vocabulary
in Eq.~(\ref{eq.taylor.quant}), i.e. $\sigma(M) \approx 0.1 \mu(M)$, we can calculate the
z-score of the observation $(N,M)$ as 
\begin{equation}\label{eq.appl.z2}
 z_{(N,M)} \approx \frac{N - \mu(M)}{0.1 \mu(M)} = 10 \left(\frac{N}{\mu(M)} - 1 \right),
\end{equation}
in which we need to include the expected vocabulary growth, $\mu(M)$, from a given generative model (e.g., Heaps' law with two scalings~\cite{Gerlach2012}).
We can now: 
i) for a single text $(N,M)$, assign a value of lexical richness, the z-score $z_{(N,M)}$, taking into account deviations from the pure Heaps' law which should be included in $\mu(M)$;
ii) given two texts $(N_1,M_1)$ and $(N_2,M_2)$, compare directly the respective z-scores $z_{(N_1,M_1)}$ and $z_{(N_2,M_2)}$ in order to assess which text has a higher lexical richness independent of the difference in the textlengths; and
iii) estimate the statistical significance of the difference in vocabulary by considering $\Delta z := z_{(N_1,M_1)} - z_{(N_2,M_2)}$, which is distributed according to $\Delta z \overset{d}{\sim} \mathcal{N}(0,2)$ since $z \overset{d}{\sim} \mathcal{N}(0,1)$.
Point (iii) implies that the difference in the vocabulary richness of two texts is statistically significant on a $95\%$-confidence level if $|\Delta z | >2.77$, i.e. in this case there is at most a $5\%$ chance that the observed difference originates from topic fluctuations.
As a rule of thumb, for two texts of approximately the same length ($N(M)\approx \mu(M)$), the relative difference in the vocabulary has to be larger than $27.7\%$ in order to be sure on a $95\%$-confidence level that the difference is not due to expected topic fluctuations.

\begin{figure*}[!bt]
\centering
\includegraphics[width=1.0\columnwidth]{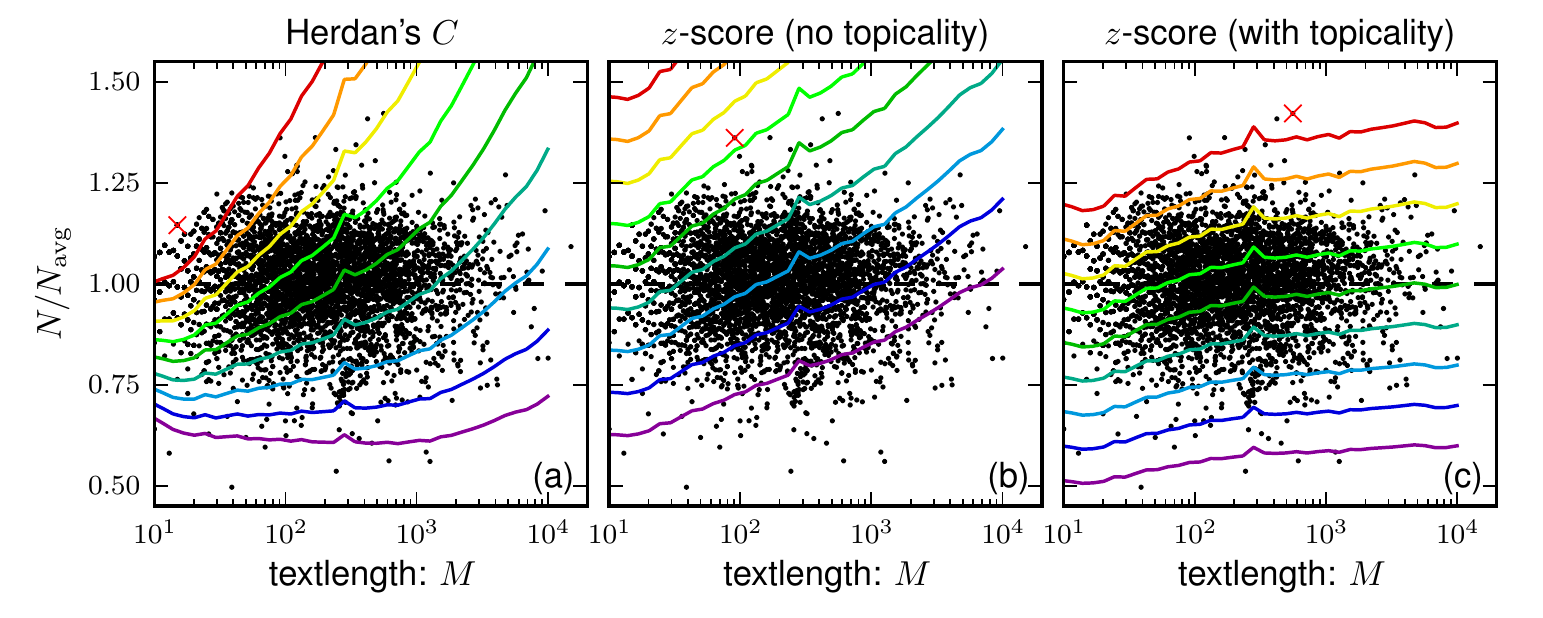}
\caption{Measures of vocabulary richness. For $5000$ randomly selected articles from the Wikipedia database (black dots), we compute the ratio between the number of different words $N(M)$ and the average number of different words $N_{\mathrm{avg}}(M)$  (empirically determined from all articles with the same textlength $M$).
We compare the predictions of different measures of vocabulary  richness (solid lines): (a) Herdan's $C$ and (b+c) $z$-score, Eq.~(\ref{eq.appl.z2}), in which we calculate the expected null model, $\mu(M)$, according to Eq.~(\ref{eq.app.ze.gamma}) with parameters $\gamma=1.77$, $\tilde{r}=7830$~\cite{Gerlach2012}, and $a\rightarrow \infty$ (in b) or $a=0.08$ (in c).
The solid lines are contours corresponding to values of $N(M)$ that yield the same measure of vocabulary richness varying from rich (red: $C=0.98$ and $z=4$) to poor (purple: $C=0.8$ and $z=-4$) vocabulary. 
The article with the richest vocabulary according to each measure is marked by $\times$ (red).
}
\label{fig.4}
\end{figure*}

We illustrate this approach for the vocabulary richness of Wikipedia articles.
As a proxy for the true vocabulary richness, we measure how much the vocabulary of each article, $N(M)$, exceeds the average vocabulary $N_{\mathrm{avg}}(M)$ with the same textlength $M$ empirically determined from all articles in the Wikipedia. In practice however, when assessing the vocabulary richness of a single article, information of $N_{\mathrm{avg}}(M)$ from an ensemble of texts is usually not available and measures such as the ones described above are needed.
In Fig.~\ref{fig.4} we compare the accuracy of measures of vocabulary richness according to Herdan's $C$, Fig.~\ref{fig.4}(a), and the $z$-score, Fig.~\ref{fig.4}(b+c). 
For the latter, we use Eq.~(\ref{eq.appl.z2}) and calculate $\mu(M)$ from Poisson word usage by fixing Zipf's law and assuming Gamma-distributed word-frequencies across documents, see \ref{sec.app.gamma} for details.
We see in Fig.~\ref{fig.4}(a) that Herdan's $C$ shows a strong bias towards assigning high values of $C$ to shorter texts: following a line with constant $C$ we observe for $M\gtrsim 10$ articles with a vocabulary below average while for $M>1000$ articles with a vocabulary above average.
A similar (weaker) bias is observed in Fig.~\ref{fig.4}(b) for the calculation of the $z$-score for the case in which we consider deviations from the pure Heaps' law but treat frequencies of individual words as fixed, i.e. ignoring topicality.
The $z$-score calculations including topicality in Fig.~\ref{fig.4}(c) show that we obtain a measure of vocabulary richness which is approximately unbiased with respect to the textlength $M$ (contour lines are roughly horizontal).
Furthermore, in contrast to the two other measures, we correctly assign the highest $z$-score to the article with the highest ratio $N(M)/N_{\mathrm{avg}}(M)$.
Altogether, this implies that it is not only important to take into account deviations from the pure Heaps' law but that it is crucial to consider topicality in the form of a quenched average.

\section{Discussion}\label{sec.discussion}
In summary, we used large text databases to investigate the scaling between vocabulary size $N$ (number of different words) and database size $M$. Besides the usual analysis of the average vocabulary size (Heaps' law), we measured the standard deviation across different texts with the same length $M$. We found that the relative fluctuations (standard deviation divided by the mean) do not decay with $M$, in contrast to simple sampling processes. 
We explained this observation using a simple stochastic process (Poisson usage of words) in which we account for topical aspects of written text, i.e. the frequency of an individual word is not treated as fixed across different documents. This heterogeneous dissemination of words across different texts leads to a reduction of the expected size of the vocabulary and to an increase in the variance.
We have further shown the implications of these findings by proposing a practical measure of vocabulary richness which allows for a comparison of the vocabulary of texts with different lengths, including the quantification of statistical significance.

Our finding of anomalous fluctuation scaling implies that the vocabulary is a non-self-averaging quantity, meaning that the vocabulary of a single text is not representative of the whole ensemble. Here we emphasized that topicality can be responsible for this effect. While the existence of different topics is obvious for a collection of articles as broad in content as the Wikipedia, our analysis shows that we can apply the same reasoning for the Google-ngram data, in which case the frequency variation is measured at different times. This offers a new perspective on language change~\cite{Baronchelli2012}: the difference in the vocabulary from different years can be seen as a shift in the topical content over time. Similarly, other systematic fluctuations (e.g., across different authors or in the parameters of the Zipf's law) can play a similar role as topicality.

Beyond linguistic applications, allometric scaling~\cite{West1997,Bettencourt2007a} and other sublinear scalings similar to Heaps' law~\cite{Brainerd1982,GarciaMartin2006,Cattuto2009,Krapivsky2013,Perotti2013,Tria2013}   have been observed in different complex systems. Our results show the importance of studying fluctuations around these scalings and provide a theoretical framework for the analysis.   

\section*{Acknowledgements}
  
We thank Diego Rybski for insightful discussion on fluctuation scaling.


\appendix
\section{Data}\label{sec.app.data}
The Wikipedia database consists of the plain text of all $3,743,306$ articles from a snapshot of the complete English Wikipedia~\cite{Wikimedia}. The PlosOne database consists of all $76,723$ articles published in the journal PlosOne which were accessible via the API at the time of the data collection~\cite{Plosone}. The Google-ngram database is a collection of printed books counting the number of times a word appears in a given year $t \in [1520-2008]$~\cite{mish+11}. We treat the collection of all books published in the the same year as a single document, yielding $393$ observations for different $t$.

We apply the same filtering for each database: i) we decapitalize each word (e.g. ``the" and ``The" are counted as the same word) and ii) we restrict ourselves to words consisting uniquely of letters present in the alphabet of the English language. This is meant as a conservative approach in order to minimize the influence of foreign words, numbers (e.g. prices), or scanning problems which are present in the raw data (for details on the preprocessing see~\cite{Gerlach2012}).

Due to peculiarities of the individual databases the data (Data: $\mu,\sigma$) in Fig.~\ref{fig.1}, i.e. the calculation of the curves $\mu(M)$ and $\sigma(M)$ conditioned on the textlength $M$,  is constructed in a slightly different way in each case. 
In the Wikipedia data we order all datapoints $N(M)$ (of the full article) according to textlength $M$ and consider $1000$ consecutive datapoints (in $M$) from which we calculate the average value of the textlength $M$, and the conditional mean, $\mu(M)$, and variance, $\sigma(M)$, of the vocabulary $N$.
In the PlosOne data the length of all articles is much more concentrated, which is why we consider the full trajectory $N(M)$ with $M=1,2,...,M_{\mathrm{max}}$ for each individual article.
For an arbitrary value of $M$ we calculate $\mu(M)$ and $\sigma(M)$ from the ensemble of all articles with vocabulary $N$ at the particular textlength $M$.
In the Google-ngram data we impose a logarithmic binning in $M$ such that we can calculate $\mu(M)$ and $\sigma(M)$ from a finite number of samples in each bin. 

\section{Poisson Null Model}\label{sec.app.poissonnull}
The number of different words in each realization of the Poisson process is given by
\begin{equation}
 N(M) = \sum_r I\left[ n_r (M, F_r)  \right],
\end{equation}
in which $n_r$ is the integer number of times the word $r$ occurs in a Poisson process of length $M$ with frequency $F_r$ and $I[x]$ is an indicator-type function, i.e. $I[x=0]=0$ and $I[x\ge1]=1$.
Averaging over realizations of the Poisson process requires the calculation of $\mathbb{E}[I[n_r(M, F_r)]] \equiv \langle I[n_r(M)] \rangle = 1-e^{-M F_r}$, which is the probability that the word with rank $r$ appears at least once in a text of length $M$. 
Considering all words we obtain
\begin{eqnarray}
  \mathbb{E} \left[ N(M) \right]  &=  \sum_r \langle I[n_r(M)] \rangle = \sum_r 1- e^{-M F_r} , \\
  \mathbb{V} \left[ N(M) \right] &\equiv  \mathbb{E} \left[ N(M)^2 \right] -\mathbb{E} \left[ N(M) \right]^2\\
   &= \sum_{r,r'} \langle I \left[n_r\right] I\left[n_{r'} \right] \rangle - \sum_{r,r'} \langle I \left[n_r\right] \rangle \langle I\left[n_{r'} \right] \rangle \\
   &= \sum_r \langle I \left[n_r\right]^2 \rangle + \sum_{\underset{r \neq r'}{r,r'}}  \langle I \left[n_r\right]  I\left[n_{r'} \right] \rangle- \sum_{r,r'} \langle I \left[n_r\right] \rangle \langle I\left[n_{r'} \right] \rangle\\
   &= \sum_r \langle I \left[n_r\right] \rangle + \sum_{\underset{r \neq r'}{r,r'}}  \langle I \left[n_r\right] \rangle \langle I\left[n_{r'} \right] \rangle- \sum_{r,r'} \langle I \left[n_r\right] \rangle \langle I\left[n_{r'} \right] \rangle\\
   &= \sum_r e^{-M F_r} - e^{-2 M F_r}
\end{eqnarray}
where we used that $I[x]^2 = I[x]$ and that Poisson processes of different words ($r \neq r'$) are independent of each other.

\section{Calculation $\mathbb{E}_q \left[ N(M)^2 \right] $}\label{sec.app.N2}
\begin{eqnarray}
  \mathbb{E}_q \left[ N(M)^2 \right] 
  &=&  \left\langle N(M)^{(i,j)} N(M)^{(i,j)}\right\rangle_{i,j}  \\
  &=&  \left\langle \sum_{r,r'} I [ n^{(i)}_{r} (M, F^{(j)}_{r})  ] I [ n^{(i)}_{r'} (M, F^{(j)}_{r'})  ] \right\rangle_{i,j} \\
  &=&  \sum_{r}  \left\langle I[ n^{(i)}_r (M, F^{(j)}_r)  ]^2  \right\rangle_{i,j} \\
      &&+  \left\langle \sum_{r}\sum_{r'\neq r} I [ n^{(i)}_r (M, F^{(j)}_r)  ] I [ n^{(i)}_{r'} (M, F^{(j)}_{r'})  ] \right\rangle_{i,j} \nonumber\\
    &=&  \sum_{r} \left\langle I[ n^{(i)}_r (M, F^{(j)}_r)  ] \right\rangle_{i,j} \\
      &&+  \sum_{r}\sum_{r'\neq r} \left\langle \langle I[ n^{(i)}_r (M, F^{(j)}_r)  ] \rangle_{i} \langle I[ n^{(i)}_{r'} (M, F^{(j)}_{r'})  ] \rangle_{i} \right\rangle_{j} \nonumber\\
  &=& \sum_r 1 - \left\langle e^{-M F_r} \right\rangle \\
  &&+ \sum_{r}\sum_{r'\neq r} \left\langle \left(1-e^{-MF_r}\right)\left(1-e^{-MF_{r'}}\right) \right\rangle \nonumber
\end{eqnarray}
where we used $I[x]^2=I[x]$, Eq.~(\ref{eq.ze.poiss}), and that two Poisson process of different words ($r \neq r'$)  with a given set of frequencies $F^{(j)}_r$ are independent of each other.

\section{Adding Texts}\label{sec.app.add}
In this section we show the calculation for the quenched averages of the mean and the variance of the vocabulary growth when considering a text of length $M'$ from the concatenation of $k$ different texts of length $M_i$ with $M' = \sum_{i=1}^k M_i$.
We will first focus on the case $k=2$, i.e. $M'=M_1+M_2$, from which we can easily generalize to arbitrary $k$.

We consider the vocabulary growth, $N(M')$, as a random variable in which we concatenate two independent realizations of the stochastic process introduced in Sec.~\ref{sec.topic.gen} indicated by subscript $(1)$ and $(2)$ respectively:
\begin{eqnarray}
 N(M'=M_1+M_2) &=& \sum_r I[ n_r^{(1)} (M_1, F_r^{(1)}) +  n_r^{(2)} (M_2, F_r^{(2)})  ]\\
      &=& \sum_r I[ n_r^{(1)} (M_1, F_r^{(1)}) ] + I[ n_r^{(2)} (M_2, F_r^{(2)}) ]\\
      &&- I[ n_r^{(1)} (M_1, F_r^{(1)}) ] I[  n_r^{(2)} (M_2, F_r^{(2)})  ] \nonumber
\end{eqnarray}
in which the word $r$ is counted as part of the vocabulary if it appears in either of the two concatenated realizations of the stochastic process.
In the same spirit as in Sec.~\ref{sec.topic.gen}, taking expectation values requires averaging over all realizations  of the Poisson process ($i_1,i_2$) given the frequencies $F_r^{(j_1)},F_r^{(j_2)}$ as well as averaging over all realizations of those frequencies ($j_1,j_2$), which we denote by $\left\langle \cdot \right\rangle_{i_1,i_2,j_1,j_2}$.
For the individual terms appearing in $N(M'=M_1+M_2)$  we get 
\begin{eqnarray}
\left\langle I[ n_r^{(i_1)} (M_1, F_r^{(j_1)}) ] \right\rangle_{i_1,i_2,j_1,j_2} &=& 1 - \langle e^{-M_1 F_r^{(j_1)}} \rangle_{j_1} \\
\left\langle I[ n_r^{(i_2)} (M_2, F_r^{(j_2)}) ] \right\rangle_{i_1,i_2,j_1,j_2} &=& 1 - \langle e^{-M_2 F_r^{(j_2)}} \rangle_{j_2},
\end{eqnarray}
\begin{eqnarray}
\langle I\left[ n_r^{(i_1)} (M_1, F_r^{(j_1)}) \right] & I\left[  n_{r'}^{(i_2)} (M_2, F_{r'}^{(j_2)})  \right] \rangle_{i_1,i_2,j_1,j_2} =\nonumber \\ &\left(1 - \langle e^{-M_1 F_{r}^{(j_1)}} \rangle_{j_1}\right) \left( 1 - \langle e^{-M_2 F_{r'}^{(j_2)}} \rangle_{j_2} \right),
\end{eqnarray}
in which we can separate the average over $(i_1,j_1)$ and $(i_2,j_2)$ assuming that the two concatenated realizations $(i_1,j_1)$ and $(i_2,j_2)$ of the original stochastic process are independent. 
For the calculation of the expectation of $N(M'=M_1+M_2)^2$ we get higher order terms for $r \neq r'$:
\begin{eqnarray}
\langle I\left[ n_r^{(i_1)} (M_1, F_r^{(j_1)}) \right] & I\left[  n_{r'}^{(i_1)} (M_1, F_{r'}^{(j_1)})  \right] \rangle_{i_1,i_2,j_1,j_2} =\nonumber \\ & \langle \left(1 -  e^{-M_1 F_{r}^{(j_1)}} \right) \left( 1 -  e^{-M_1 F_{r'}^{(j_1)}}\right) \rangle_{j_1}.  
\end{eqnarray}
From this we can evaluate the mean and variance
\begin{eqnarray}
 \mathbb{E}_q[N(M'=M_1+M_2)] &=& \sum_r 1 - \left\langle e^{-M_1 F_r} \right\rangle \left\langle e^{-M_2 F_r} \right\rangle \\
  \mathbb{V}_q[N(M'=M_1+M_2)] &=& \sum_r \left\langle e^{-M_1 F_r} \right\rangle \left\langle e^{-M_2 F_r} \right\rangle - \left\langle e^{-2 M_1 F_r} \right\rangle \left\langle e^{-2 M_2 F_r} \right\rangle \nonumber\\
&&+ \sum_{r,r'} \left\langle e^{-M_1 F_r} e^{-M_1 F_{r'}}\right\rangle \left\langle e^{-M_2 F_r} e^{-M_2 F_{r'}}\right\rangle \\
&&-\left\langle e^{-M_1 F_r}\right\rangle \left\langle e^{-M_1 F_{r'}}\right\rangle \left\langle e^{-M_2 F_r}\right\rangle \left\langle e^{-M_2 F_{r'}}\right\rangle.\nonumber
\end{eqnarray}

Generalizing to the concatenation of an arbitrary number of $k$ texts can be treated in the very same way, however, we will only state the result for the case of adding $k$ texts of equal length $M$ such that $M'=k \times M$:
\begin{eqnarray}
 \mathbb{E}_q[N(M'=k\times M)] &=& \sum_r 1 - \left\langle e^{-M F_r} \right\rangle^k  \\
  \mathbb{V}_q[N(M'=k \times M)] &=& \sum_r \left\langle e^{-M F_r} \right\rangle^k -  \left\langle e^{-2M F_r} \right\rangle^k \\
&&+ \sum_{r,r'} \left\langle e^{-M F_r} e^{-M F_{r'}} \right\rangle^k - \left\langle e^{-M F_r} \right\rangle^k \left\langle e^{-M F_{r'}} \right\rangle^k \nonumber.
\end{eqnarray}

\section{Vocabulary Growth for Gamma-distributed frequency and a double power-law}\label{sec.app.gamma}
Assuming a Gamma-distribution for the distribution of the frequency of single words across different texts~\cite{Church1995}
\begin{equation}
P_{\Gamma}(F_r = x;a,b) = \frac{1}{\Gamma(a)} b^{-a} x^{a-1}e^{-x/b}
\end{equation}
we can calculate the quenched average
\begin{equation}
 \left\langle e^{-MF_r} \right\rangle = \int \mathrm{d}x P_{\Gamma}(F_r = x;a,b) e^{-Mx} = (1+bM)^{-a}.
\end{equation}
If we assume that the distribution of frequencies for all words is given by the same shape-parameter $a$ (e.g. $a=1$ corresponds to an exponential distribution) and fix the mean of the distribution, given by $\left\langle F_r \right\rangle=ab$ we get $\left\langle e^{-MF_r} \right\rangle = (1+  M \left\langle F_r \right\rangle /a)^{-a}$.
Assuming a double power-law for the average rank-frequency distribution~\cite{Gerlach2012} with parameters $\gamma$ and $\tilde{r}$, i.e. $\left\langle F_r \right\rangle =C r^{-1}$ for $r\leq \tilde{r}$ and $\left\langle F_r \right\rangle =C \tilde{r}^{\gamma-1}r^{-\gamma}$ for $r> \tilde{r}$, where $C=C(\tilde{r},\gamma)$ is the normalization constant determined by imposing $\sum_r \left\langle F_r \right\rangle =1$,
we can calculate the vocabulary growth according to Eq.~(\ref{eq.ze.mean}) analytically in the continuum approximation by substituting $x:=\left\langle F_r \right\rangle$:
\begin{eqnarray}
  \mathbb{E}_q \left[ N(M) \right] &=& \sum_r 1- (1+  M \left\langle F_r \right\rangle /a)^{-a} \\
   &=&  - \int_0^1 \mathrm{d}x \frac{\mathrm{d}r}{\mathrm{d}x}\left[1- (1+  M x /a)^{-a}\right]
\end{eqnarray}
which can be expressed in terms of the ordinary hypergeometric function $H:={}_{2}F_{1}$~\cite{Abramowitz1972} yielding
\begin{eqnarray}\label{eq.app.ze.gamma}
 \mathbb{E}_q \left[ N(M) \right]&=& \tilde{r}-C + \tilde{r} \left[ H(a,-\frac{1}{\gamma},1-\frac{1}{\gamma},-\frac{CM}{a\tilde{r}})-1 \right]\nonumber\\
 &-&C\left(1+\frac{M}{a}\right)^{-a}\left[a\frac{\Gamma(1+a)}{\Gamma(2+a)}H(1,1,2+a,-\frac{a}{M})-1 \right]\\
 &+&\tilde{r}\left(1+\frac{CM}{a\tilde{r}}\right)^{-a}\left[a\frac{\Gamma(1+a)}{\Gamma(2+a)}H(1,1,2+a,-\frac{a\tilde{r}}{CM})-1 \right], \nonumber
\end{eqnarray}
where the vocabulary growth $\mathbb{E}_q \left[ N(M) \right]$ is parametrized by $\gamma$, $\tilde{r}$, and $a$.

In the limit $a\rightarrow \infty$ the Gamma distribution $P_{\Gamma}(F_r = x;a,b)$ with given mean $\left\langle F_r \right\rangle = ab = \mathrm{const.}$ converges to a Gaussian with $\sigma^2=\left\langle F_r \right\rangle^2/a$. For $a\rightarrow \infty$, $\sigma^2 \rightarrow 0$ and we recover the Poisson null model, Eqs.~(\ref{eq.ze.mean},\ref{eq.ze.sdev}), in which the individual frequencies $F_r$ are fixed (annealed average).


\providecommand{\newblock}{}

\end{document}